\documentstyle[NATO,epsf]{CRCKAPB}

\begin{opening}

\title{Double Degenerates from the Supernova Ia Progenitor Survey (SPY)}

\author{C. Karl}
\author{R. Napiwotzki}
\author{U. Heber}
\author{T. Lisker}
\institute{Dr. Remeis Sternwarte, Astronomisches Institut der
 Universit\"at Erlangen~-~N\"urnberg, Sternwartstr. 7, 96049 Bamberg, Germany}
\author{G. Nelemans}
\institute{Institute of Astronomy, Madingley Road, CB3 0HA, Cambridge, UK}
\author{N. Christlieb}
\author{D. Reimers}
\institute{Hamburger Sternwarte, Universit\"at Hamburg, Gojenbergsweg 112, 
 21029 Hamburg, Germany}

\end{opening}

\begin{document}

\section{Introduction}

Close binary systems consisting of two white dwarfs (double
degenerates~--~DDs) are considered possible progenitors of
supernovae type Ia. The SPY project (Napiwotzki {\em et al.},
these proceedings) surveys a large sample of white dwarfs for
radial velocity (RV) variations. DDs are the result of
close binary evolution, but the theoretical modeling of this evolution
depends heavily on some
not well known parameters \cite{N01}. Therefore, an important way to
calibrate these parameters is by comparison of observed system parameters
such as periods and mass ratios of DDs with results of 
theoretical calculations.
We present follow-up observations of four DDs found by the SPY project,
which allowed us to derive radial velocity curves
and system parameters.
Two systems are of particular interest because they have
double lined spectra.

\section{Observations}

Follow up observations carried out with
the TWIN spectrograph at the 3.5m telescope of the Calar Alto Observatory,
Spain (spectral resolution 1.0 \AA) are combined with 
the discovery spectra taken with the UVES
echelle spectrograph at the ESO UT2 telescope at Paranal, Chile
(spectral resolution 0.3 \AA).

\section{Data analysis and results}
\begin{figure}
\epsfxsize=13cm
\epsfysize=11cm
\epsfbox{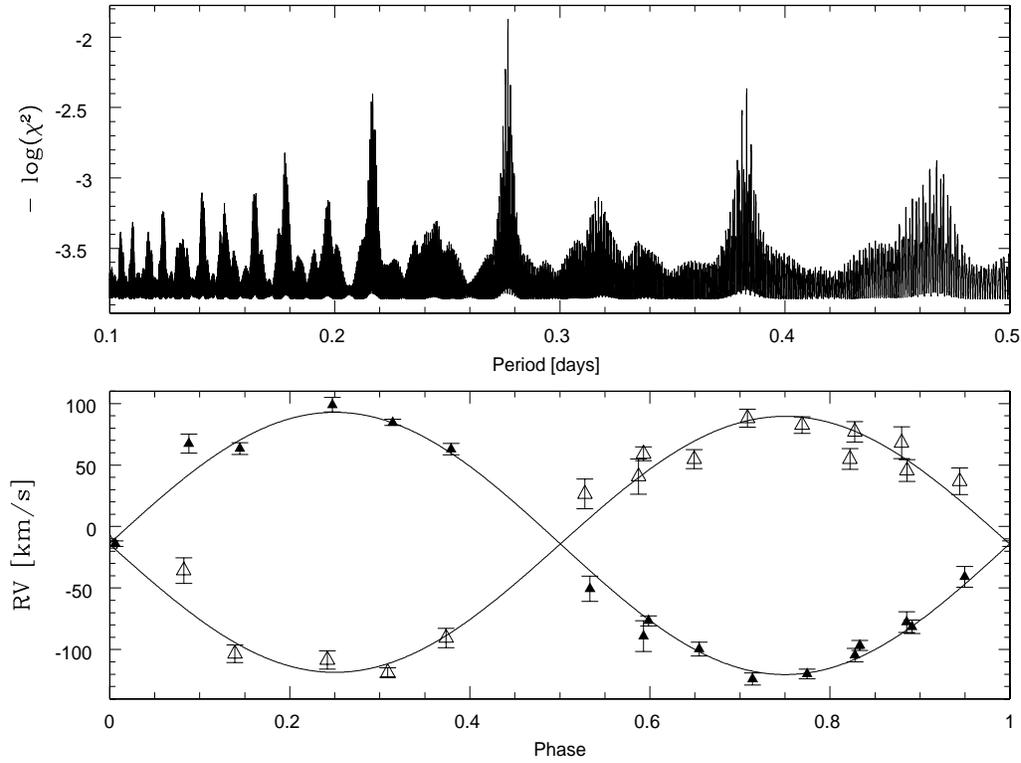}
\caption{Upper part: Combined power spectrum of the double lined system
HE~2209$+$1444. Lower part: Radial velocity curve. The brighter component
(filled triangles) is as well plotted as the fainter one (open triangels).}
\label{he2209}
\end{figure}

For the analysis of our programme stars, we first computed the RV curves.
We measured the RV of each spectrum by fitting the observed
H${\alpha}$ line profile with a Gaussian for the line core and a Lorentzian
for the wing plus a linear continuum.
With respect to the center of the Gaussian we determined the RV.
By fitting sinosoidal functions to a range
of periods we obtained {\em power spectra}. These spectra
indicate the quality of the sine~-~fits as a function
of period.
{\em Combined power spectra} were produced for the double lined systems
by simply adding the
$\chi^2$ values of the two {\em individual power spectra}.
In this way we got the most probable period of the complete system. 
Then, having fixed the period, we fitted the semi-amplitudes,
the system velocities and the ephemeries using sine curves again.
Results for the system HE~2209$-$1444 are displayed in figure \ref{he2209}. 

\paragraph{\bf HE~2209$-$1444:} This is the first double lined system
we analyzed. From the {\em combined power spectrum} we derived a period
of 0.2769~days (see figure \ref{he2209} and the following ephemeris).
The Ephemeris for the time T$_0$ (defined as the conjunction time
at which star A moves from the red to the blue side of the RV curve) is
\begin{eqnarray} {\rm Hel.JD(T}_0) = 2,452,096.9010 + 0.2769 \times E .
\end{eqnarray}
The periods derived from both {\em individual power spectra} agree within
0.8~sec. Results are provided in table \ref{orbital parameters}.
Also given are the semi amplitudes $K_{\rm A/B}$ and the system velocities
$\gamma_0$. As can be seen, the ratio of the semi amplitudes is
near unity, which means that the two WDs must have similar masses.
Fitting observed Balmer line profiles simultaneously with a grid
of synthetic spectra from NLTE model atmospheres we derive an
effective temperature of 8450 resp. 7700 K and $\log g = 8.07$
for both components.
By comparism with evolutionary calculations \cite{B97}
we derived masses of 0.63 $M_{\odot}$.
We computed the merging time due to gravitational wave
radiation to be 4.2 Gyr for this system, i.e. well below a
Hubble time. The total mass is remarkably close to the
Chandrasekhar limit, only 10\% below it.

\begin{table}[htb]
\begin{center}
\caption{Orbital parameters of the program stars}
\label{orbital parameters}
\begin{tabular}{llccc} \hline
System        & { } & $\gamma_0$ [km/s] & K [km/s]  \\ \hline
HE 2209 - 1444& A   & -12.2            & 108.2 \\
{ }           & B   & -13.5            & 103.3 \\
WD 1349 + 144 & A   & -18.2            & 74.5 \\
{ }           & B   & -10.3            & 66.9 \\
WD 1824 + 040 & { } & 48.5             & 59.5 \\
EGB5          & { } & 69.6             & 16.1 \\ \hline
\end{tabular}
\end{center}
\end{table}

\paragraph{\bf WD~1349$+$144:}
This is another double~-~lined system.
Results from the single star anlysis are given in table
\ref{orbital parameters}. The {\em combined power spectrum} shows
a peak at a period of 2.2094 days, but some aliases with
slightly different periods cannot be ruled out. The ephemeris of
this system is
\begin{eqnarray}   {\rm Hel.JD(T}_0) = 2,452,096.7350 + 2.2094 \times E .
\end{eqnarray}
The Balmer line spectra of both components are very similar and
the mass ratio is very close to unity, thus we assumed identical
stellar parameters for both stars. Fitting a spectrum taken near
conjunction we derived an effective temperature of
$T_{\rm eff} \approx 16.600$ K and $\log g \approx$ 7.65.
The resulting masses are 0.44 $M_{\odot}$ for each component,
which leads to a merging time of 2000 Gyr,
much larger than a Hubbe time.

\paragraph{\bf WD~1824$+$040:}
The spectrum of this 
well known RV variable DA~WD 
is single lined. The period derived from the {\em power spectrum}
is 6.2663 days which is in perfect agreement with previous results
from \citeauthor{MM99} (\citeyear{MM99}). The epehmeris is
\begin{eqnarray}   {\rm Hel.JD(T}_0) = 2,451979.1918 + 6.2663 \times E .
\end{eqnarray}
From the fundamental parameters $T_{\rm eff} = 14795$K and $\log g = 7.61$
\cite{B95} we computed a mass of 0.44 $M_{\odot}$ for the visible
component.
If we adopt the most probable inclination angle of 54$^{\rm o}$,
we derive a mass of 0.63 $M_{\odot}$ for the invisible component
from the mass function.
This system will merge within 24000 Gyr, again much larger than
the Hubble time.

\paragraph{\bf EGB~5:}
EGB~5 is a centrel star of an old planetary nebula (PN~G211.9$+$22.6).
In addition to the H${\alpha}$ and H${\beta}$ lines
He{\sc i}(4471\AA{})
and He{\sc ii}(4686\AA{}) were available for RV measurements.
From the {\em power spectrum}
we found a strong peak at 1.1806 days, but there is a slightly weaker
peak at 0.5505 days. Inspection of the RV curves for both periods cannot
clearly favour one period over the other.

\paragraph{Acknowledgements.} We express our gratitude to the ESO staff
for conducting the service observations and pipeline reductions.
We gratefully acknowledge the assistance of the Calar Alto staff.


\begin{thebibliography}{}
\bibitem[\protect\citeauthoryear{Bl\"ocker {\it et al.}}{1997}]{B97}
Bl\"ocker, T., Herwig, F., Driebe, T., Bramkamp, H., Sch\"onberner D., 1997,
{\it proceedings of the 10th European Workshop on White Dwarfs}, {\bf 57} 
\bibitem[\protect\citeauthoryear{Bragaglia {\it et al.}}{1995}]{B95}
Bragaglia, A., Renzini, A., Bergeron, P., 1995, {\it ApJ}, {\bf 433}, 735
\bibitem[\protect\citeauthoryear{Maxted and Marsh}{1999}]{MM99}
Maxted, P.F.L., Marsh, T.R., 1997, {\it MNRAS}, {\bf 307}, 122
\bibitem[\protect\citeauthoryear{Nelemans {\it et al.}}{2001}]{N01}
Nelemans, G., Yungelson, L.R., Portegis Zwart, S.F. Verbunt, F., 2001,
{\it A\& A}, {\bf 365}, 491

\end{thebibliography}
\end{document}